\font\ninerm=cmr9
\font\sevenrm=cmr7
\font\sixrm=cmr6
\font\fiverm=cmr5
\font\ninei=cmmi9
\font\sixi=cmmi6
\font\fivei=cmmi6
\font\ninesy=cmsy9
\font\sixsy=cmsy6
\font\fivesy=cmsy5
\font\tenex=cmex10
\font\nineit=cmti9
\font\ninesl=cmsl9
\font\ninett=cmtt9
\font\ninebf=cmbx9
\font\sixbf=cmbx6
\font\fivebf=cmbx5

\def\ninepoint{\def\rm{\fam0\ninerm}
  \textfont0=\ninerm \scriptfont0=\sixrm \scriptscriptfont0=\fiverm
  \textfont1=\ninei \scriptfont1=\sixi \scriptscriptfont0=\fivei
  \textfont2=\ninesy \scriptfont2=\sixsy \scriptscriptfont2=\fivesy
  \textfont3=\tenex \scriptfont3=\tenex \scriptscriptfont3=\tenex
  \textfont\itfam=\nineit  \def\it{\fam\itfam\nineit}%
  \textfont\slfam=\ninesl  \def\sl{\fam\slfam\ninesl}%
  \textfont\ttfam=\ninett  \def\tt{\fam\ttfam\ninett}%
  \textfont\bffam=\ninebf  \scriptfont\bffam=\sixbf
   \scriptscriptfont\bffam=\fivebf  \def\bf{\fam\bffam\ninebf}%
  \normalbaselineskip=11pt
  \setbox\strutbox=\hbox{\vrule height8pt depth3pt width0pt}%
  \let\sc=\sevenrm  \normalbaselines\rm}

\font\eightrm=cmr8
\font\sevenrm=cmr7
\font\sixrm=cmr6
\font\fiverm=cmr5
\font\eighti=cmmi8
\font\sixi=cmmi6
\font\fivei=cmmi6
\font\eightsy=cmsy8
\font\sixsy=cmsy6
\font\fivesy=cmsy5
\font\tenex=cmex10
\font\eightit=cmti8
\font\eightsl=cmsl8
\font\eighttt=cmtt8
\font\eightbf=cmbx8
\font\sixbf=cmbx6
\font\fivebf=cmbx5

\def\eightpoint{\def\rm{\fam0\eightrm}
  \textfont0=\eightrm \scriptfont0=\sixrm \scriptscriptfont0=\fiverm
  \textfont1=\eighti \scriptfont1=\sixi \scriptscriptfont0=\fivei
  \textfont2=\eightsy \scriptfont2=\sixsy \scriptscriptfont2=\fivesy
  \textfont3=\tenex \scriptfont3=\tenex \scriptscriptfont3=\tenex
  \textfont\itfam=\eightit  \def\it{\fam\itfam\eightit}%
  \textfont\slfam=\eightsl  \def\sl{\fam\slfam\eightsl}%
  \textfont\ttfam=\eighttt  \def\tt{\fam\ttfam\eighttt}%
  \textfont\bffam=\eightbf  \scriptfont\bffam=\sixbf
   \scriptscriptfont\bffam=\fivebf  \def\bf{\fam\bffam\eightbf}%
  \normalbaselineskip=9pt
  \setbox\strutbox=\hbox{\vrule height7pt depth2pt width0pt}%
  \let\sc=\sixrm  \normalbaselines\rm}

\catcode `\!=11
\catcode `\@=11

\let\!tacr=\\ 

\newdimen\LineThicknessUnit 
\newdimen\StrutUnit            
\newskip \InterColumnSpaceUnit  
\newdimen\ColumnWidthUnit     
\newdimen\KernUnit

\let\!taLTU=\LineThicknessUnit 
\let\!taCWU=\ColumnWidthUnit   
\let\!taKU =\KernUnit          

\newtoks\NormalTLTU
\newtoks\NormalTSU
\newtoks\NormalTICSU
\newtoks\NormalTCWU
\newtoks\NormalTKU

\NormalTLTU={1in \divide \LineThicknessUnit by 300 }
\NormalTSU ={\normalbaselineskip
  \divide \StrutUnit by 11 }  
\NormalTICSU={.5em plus 1fil minus .25em}  
\NormalTCWU ={.5em}
\NormalTKU  ={.5em}

\def\NormalTableUnits{%
  \LineThicknessUnit   =\the\NormalTLTU
  \StrutUnit           =\the\NormalTSU
  \InterColumnSpaceUnit=\the\NormalTICSU
  \ColumnWidthUnit     =\the\NormalTCWU
  \KernUnit            =\the\NormalTKU}
 
\NormalTableUnits

\newcount\LineThicknessFactor    
\newcount\StrutHeightFactor      
\newcount\StrutDepthFactor       
\newcount\InterColumnSpaceFactor 
\newcount\ColumnWidthFactor      
\newcount\KernFactor
\newcount\VspaceFactor

\newcount\TracingKeys 
\newcount\TracingFormats  

\def\BeginTableParBox#1{%
  \vtop\bgroup 
    \hsize=#1
    \normalbaselines 
    \let~=\!ttTie
    \let\-=\!ttDH
    \the\EveryTableParBox} 
  
\def\EndTableParBox{%
    \MakeStrut{0pt}{\StrutDepthFactor\StrutUnit}
  \egroup} 

\newtoks\EveryTableParBox
\EveryTableParBox={%
  \parindent=0pt
  \raggedright
  \rightskip=0pt plus 4em 
  \relax}

\newtoks\EveryTable
\newtoks\!taTableSpread

\newskip\LeftTabskip
\newskip\RightTabskip

\newcount\!taCountA
\newcount\!taColumnNumber
\newcount\!taRecursionLevel 

\newdimen\!taDimenA  
\newdimen\!taDimenB  
\newdimen\!taDimenC  
\newdimen\!taMinimumColumnWidth

\newtoks\!taToksA

\newtoks\!taPreamble
\newtoks\!taDataColumnTemplate
\newtoks\!taRuleColumnTemplate
\newtoks\!taOldRuleColumnTemplate
\newtoks\!taLeftGlue
\newtoks\!taRightGlue

\newskip\!taLastRegularTabskip

\newif\if!taDigit
\newif\if!taBeginFormat
\newif\if!taOnceOnlyTabskip

\def\TaBlE{%
  T\kern-.27em\lower.5ex\hbox{A}\kern-.18em B\kern-.1em
    \lower.5ex\hbox{L}\kern-.075em E}

{\catcode`\|=13 \catcode`\"=13
  \gdef\ActivateBarAndQuote{%
    \ifnum \catcode`\|=13
    \else
      \catcode`\|=13
      \def|{%
        \ifmmode
          \vert
        \else
          \char`\|
        \fi}%
    \fi
    \ifnum \catcode`\"=13
    \else
      \catcode`\"=13
      \def"{\char`\"}%
    \fi}}
 
{\catcode `\|=12 \catcode `\"=12 

}

\def\!thMessage#1{\immediate\write16{#1}\ignorespaces}
 
\let\!thx=\expandafter

\def\!thGobble#1{} 

\def\\{\let\!thSpaceToken= }\\ 

\def\!thHeight{height}
\def\!thDepth{depth}
\def\!thWidth{width}

\def\!thToksEdef#1=#2{%
  \edef\!ttemp{#2}%
  #1\!thx{\!ttemp}%
  \ignorespaces}

\def\!thStoreErrorMsg#1#2{%
  \toks0 =\!thx{\csname #2\endcsname}%
  \edef#1{\the\toks0 }}

\def\!thReadErrorMsg#1{%
  \!thx\!thx\!thx\!thGobble\!thx\string #1}

\def\!thError#1#2{%
  \begingroup
    \newlinechar=`\^^J%
    \edef\!ttemp{#2}%
    \errhelp=\!thx{\!ttemp}%
    \!thMessage{%
      ^^J\!thReadErrorMsg\!thErrorMsgA 
      ^^J\!thReadErrorMsg\!thErrorMsgB}%
    \errmessage{#1}%
  \endgroup}

\!thStoreErrorMsg\!thErrorMsgA{%
  TABLE error; see manual for explanation.}
\!thStoreErrorMsg\!thErrorMsgB{%
  Type \space H <return> \space for immediate help.}

\def\!thGetReplacement#1#2{%
   \begingroup
     \!thMessage{#1}
     \endlinechar=-1
     \global\read16 to#2%
   \endgroup}

\def\!thLoop#1\repeat{%
  \def\!thIterate{%
    #1%
    \!thx \!thIterate
    \fi}%
  \!thIterate 
  \let\!thIterate\relax}

\def\Smash{%
  \relax
  \ifmmode
    \expandafter\mathpalette
    \expandafter\!thDoMathVCS
  \else
    \expandafter\!thDoVCS
  \fi}
                      
\def\!thDoVCS#1{%
  \setbox\z@\hbox{#1}%
  \!thFinishVCS}
                      
\def\!thDoMathVCS#1#2{%
  \setbox\z@\hbox{$\m@th#1{#2}$}%
  \!thFinishVCS}
                      
\def\!thFinishVCS{%
  \vbox to\z@{\vss\box\z@\vss}}

\def\!thSetDimen{%
  \ifnum \!tgCode=1
    \ifx \!tgValue\empty
      \!taDimenA \StrutHeightFactor\StrutUnit
      \advance \!taDimenA \StrutDepthFactor\StrutUnit
      \divide \!taDimenA 2
    \else
      \!taDimenA \!tgValue\StrutUnit
    \fi
  \else
    \!taDimenA \!tgValue
  \fi
  \!taDimenA=\!thSign\!taDimenA\relax
  \ifmmode
    \expandafter\mathpalette
    \expandafter\!thDoMathRaise
  \else
    \expandafter\!thDoSimpleRaise
  \fi}
                      
\def\!thDoSimpleRaise#1{%
  \setbox\z@\hbox{\raise \!taDimenA\hbox{#1}}%
  \!thFinishRaise} 
                      
\def\!thDoMathRaise#1#2{%
  \setbox\z@\hbox{\raise \!taDimenA\hbox{$\m@th#1{#2}$}}%
  \!thFinishRaise}

\def\!thFinishRaise{%
  \ht\z@\z@ 
  \dp\z@\z@
  \box\z@}

\def\!thKernBack{%
  \kern -
  \ifnum \!tgCode=1 
    \ifx \!tgValue\empty 
      \the\KernFactor
    \else
      \!tgValue    
    \fi
    \KernUnit
  \else 
    \!tgValue      
  \fi
  \ignorespaces}%

\def\Vspace{%
  \noalign
  \bgroup
  \!tgGetValue\!thVspace}

\def\!thVspace{%
  \vskip
    \ifnum \!tgCode=1 
      \ifx \!tgValue\empty 
        \the\VspaceFactor
      \else
        \!tgValue    
      \fi
      \StrutUnit
    \else 
      \!tgValue      
    \fi
  \egroup} 

\def\BeginFormat{%
  \catcode`\|=12 
  \catcode`\"=12 
  \!taPreamble={}%
  \!taColumnNumber=0
  \skip0 =\InterColumnSpaceUnit
  \multiply\skip0 \InterColumnSpaceFactor
  \divide\skip0 2
  \!taRuleColumnTemplate=\!thx{%
    \!thx\tabskip\the\skip0 }%
  \!taLastRegularTabskip=\skip0 
  \!taOnceOnlyTabskipfalse
  \!taBeginFormattrue 
  \def\!tfRowOfWidths{}
  \ReadFormatKeys}

\def\!tfSetWidth{%
  \ifx \!tfRowOfWidths \empty  
    \ifnum \!taColumnNumber>0  
      \begingroup              
         \!taCountA=1          
         \aftergroup \edef \aftergroup \!tfRowOfWidths \aftergroup {%
           \aftergroup &\aftergroup \omit
           \!thLoop
             \ifnum \!taCountA<\!taColumnNumber
             \advance\!taCountA 1
             \aftergroup \!tfAOAO
           \repeat 
           \aftergroup }%
      \endgroup
    \fi
  \fi      
  \ifx [\!ttemp 
    \!thx\!tfSetWidthText
  \else
    \!thx\!tfSetWidthValue
  \fi}

\def\!tfAOAO{%
  &\omit&\omit}

\def\!tfSetWidthText [#1]{
  \def\!tfWidthText{#1}%
  \ReadFormatKeys}

\def\!tfSetWidthValue{%
  \!taMinimumColumnWidth = 
    \ifnum \!tgCode=1 
      \ifx\!tgValue\empty 
        \ColumnWidthFactor
      \else
        \!tgValue 
      \fi
      \ColumnWidthUnit
    \else
      \!tgValue 
    \fi
  \def\!tfWidthText{}
  \ReadFormatKeys}

\def\!tfSetTabskip{%
  \ifnum \!tgCode=1
    \skip0 =\InterColumnSpaceUnit
    \multiply\skip0 
      \ifx \!tgValue\empty
        \InterColumnSpaceFactor         
      \else
       \!tgValue                        
      \fi
  \else
    \skip0 =\!tgValue                   
  \fi
  \divide\skip0 by 2
  \ifnum\!taColumnNumber=0 
    \!thToksEdef\!taRuleColumnTemplate={%
      \the\!taRuleColumnTemplate 
      \tabskip \the\skip0 }
  \else
    \!thToksEdef\!taDataColumnTemplate={%
      \the\!taDataColumnTemplate 
      \tabskip \the\skip0 }
  \fi
  \if!taOnceOnlyTabskip
  \else
    \!taLastRegularTabskip=\skip0 
  \fi                             
  \ReadFormatKeys}

\def\!tfSetVrule{%
  \!thToksEdef\!taRuleColumnTemplate={%
    \noexpand\hfil
    \noexpand\vrule
    \noexpand\!thWidth
    \ifnum \!tgCode=1
      \ifx \!tgValue\empty
        \the\LineThicknessFactor      
      \else
        \!tgValue                     
      \fi
      \!taLTU                         
    \else
      \!tgValue                       
    \fi
    ####%
    \noexpand\hfil
    \the\!taRuleColumnTemplate}       
  \!tfAdjoinPriorColumn}
 
\def\!tfSetAlternateVrule{%
  \afterassignment\!tfSetAlternateA
  \toks0 =}                           

\def\!tfSetAlternateA{%
  \!thToksEdef\!taRuleColumnTemplate={%
    \the\toks0 \the\!taRuleColumnTemplate} 
  \!tfAdjoinPriorColumn}

\def\!tfAdjoinPriorColumn{%
  \ifnum \!taColumnNumber=0
    \!taPreamble=\!taRuleColumnTemplate 
    \ifnum \TracingFormats>0             
      \!tfShowRuleTemplate
    \fi
  \else
    \ifx\!tfRowOfWidths\empty  
    \else
      \!tfUpdateRowOfWidths
    \fi
    \!thToksEdef\!taDataColumnTemplate={%
      \the \!taLeftGlue
      \the \!taDataColumnTemplate
      \the \!taRightGlue}
    \ifnum \TracingFormats>0
      \!tfShowTemplates
    \fi
    \!thToksEdef\!taPreamble={%
      \the\!taPreamble
      &
      \the\!taDataColumnTemplate
      &
      \the\!taRuleColumnTemplate}
  \fi
  \advance \!taColumnNumber 1
  \if!taOnceOnlyTabskip              
    \!thToksEdef\!taDataColumnTemplate={%
       ####\tabskip \the\!taLastRegularTabskip}
  \else
    \!taDataColumnTemplate{##}%
  \fi
  \!taRuleColumnTemplate{}
  \!taLeftGlue{\hfil}
  \!taRightGlue{\hfil}%
  \!taMinimumColumnWidth=0pt
  \def\!tfWidthText{}%
  \!taOnceOnlyTabskipfalse    
  \ReadFormatKeys}

\def\!tfUpdateRowOfWidths{%
  \ifx \!tfWidthText\empty
  \else 
    \!tfComputeMinColWidth
  \fi
  \edef\!tfRowOfWidths{%
    \!tfRowOfWidths
    &%
    \omit                                  
    \ifdim \!taMinimumColumnWidth>0pt
      \hskip \the\!taMinimumColumnWidth
    \fi
    &
    \omit}}                                

\def\!tfComputeMinColWidth{%
  \setbox0 =\vbox{%
    \ialign{
       \span\the\!taDataColumnTemplate\cr
       \!tfWidthText\cr}}%
  \!taMinimumColumnWidth=\wd0 }

\def\!tfShowRuleTemplate{%
  \!thMessage{}
  \!thMessage{TABLE FORMAT}
  \!thMessage{Column: Template}
  \!thMessage{%
    \space *c: ##\tabskip \the\LeftTabskip}
  \!taOldRuleColumnTemplate=\!taRuleColumnTemplate}

\def\!tfShowTemplates{%
  \!thMessage{%
    \space \space r: \the\!taOldRuleColumnTemplate}
  \!taOldRuleColumnTemplate=\!taRuleColumnTemplate
  \!thMessage{%
    \ifnum \!taColumnNumber<10
      \space
    \fi
    \the\!taColumnNumber c: \the\!taDataColumnTemplate}
  \ifdim\!taMinimumColumnWidth>0pt
    \!thMessage{%
      \space \space w: \the\!taMinimumColumnWidth}
  \fi}

\def\!tfFinishFormat{%
  \ifnum \TracingFormats>0
    \!thMessage{%
      \space \space r: \the\!taOldRuleColumnTemplate
        \tabskip \the\RightTabskip}%
    \!thMessage{%
      \space *c: ##\tabskip 0pt}
  \fi
  \ifnum \!taColumnNumber<2
    \!thError{%
      \ifnum \!taColumnNumber=0
        No
      \else
        Only 1
      \fi
      "|"}%
      {\!thReadErrorMsg\!tfTooFewBarsA
       ^^J\!thReadErrorMsg\!tfTooFewBarsB
       ^^J\!thReadErrorMsg\!tkFixIt}%
  \fi
  \!thToksEdef\!taPreamble={%
    ####\tabskip\LeftTabskip 
    &
    \the\!taPreamble \tabskip\RightTabskip
    &
    ####\tabskip 0pt \cr}
  \ifnum \TracingFormats>1
    \!thMessage{Preamble=\the\!taPreamble}
  \fi
  \ifnum \TracingFormats>2
    \!thMessage{Row Of Widths="\!tfRowOfWidths"}
  \fi
  \!taBeginFormatfalse 
  \catcode`\|=13
  \catcode`\"=13
  \!ttDoHalign}

\!thStoreErrorMsg\!tfTooFewBarsA{%
  There must be at least 2 "|"'s (and/or "\string \|"'s)}
\!thStoreErrorMsg\!tfTooFewBarsB{%
  between \string\BeginFormat\space and \string\EndFormat\space (or ".").}

\def\ReFormat[{%
  \omit
  \!taDataColumnTemplate{##}%
  \!taLeftGlue{}%
  \!taRightGlue{}%
  \catcode`\|=12  
  \catcode`\"=12  
  \ReadFormatKeys}

\def\!tfEndReFormat{%
  \ifnum \TracingFormats>0
    \!thMessage{ReF: 
       \the\!taLeftGlue
       \hbox{\the\!taDataColumnTemplate}
       \the\!taRightGlue}
  \fi
  \catcode`\|=13
  \catcode`\"=13
  \!tfReFormat}

\def\!tfReFormat#1{%
  \the \!taLeftGlue
  \vbox{%
    \ialign{%
      \span\the\!taDataColumnTemplate\cr
       #1\cr}}%
  \the \!taRightGlue}

\def\!tgGetValue#1{%
  \def\!tgReturn{#1}
  \futurelet\!ttemp\!tgCheckForParen}

\def\!tgCheckForParen{%
  \ifx\!ttemp (%
    \!thx \!tgDoParen
  \else
    \!thx \!tgCheckForSpace
  \fi}

\def\!tgDoParen(#1){%
  \def\!tgCode{2}%
  \def\!tgValue{#1}
  \!tgReturn}

\def\!tgCheckForSpace{%
  \def\!tgCode{1}%
  \def\!tgValue{}
  \ifx\!ttemp\!thSpaceToken
    \!thx \!tgReturn        
  \else
    \!thx \!tgCheckForDigit         
  \fi}

\def\!tgCheckForDigit{%
  \!taDigitfalse
  \ifx 0\!ttemp
    \!taDigittrue
  \else
    \ifx 1\!ttemp
      \!taDigittrue
    \else
      \ifx 2\!ttemp
        \!taDigittrue
      \else
        \ifx 3\!ttemp
          \!taDigittrue
        \else
          \ifx 4\!ttemp
            \!taDigittrue
          \else
            \ifx 5\!ttemp
              \!taDigittrue
            \else
              \ifx 6\!ttemp
                \!taDigittrue
              \else
                \ifx 7\!ttemp
                  \!taDigittrue
                \else
                  \ifx 8\!ttemp
                    \!taDigittrue
                  \else
                    \ifx 9\!ttemp
                      \!taDigittrue
                    \fi
                  \fi
                \fi
              \fi
            \fi
          \fi
        \fi
      \fi
    \fi
  \fi
  \if!taDigit
    \!thx \!tgGetNumber
  \else
    \!thx \!tgReturn 
  \fi}

\def\!tgGetNumber{%
  \afterassignment\!tgGetNumberA
  \!taCountA=}
\def\!tgGetNumberA{%
  \edef\!tgValue{\the\!taCountA}%
  \!tgReturn}

\def\!tgSetUpParBox{%
  \edef\!ttemp{%
    \noexpand \ReadFormatKeys
    b{\noexpand \BeginTableParBox{%
      \ifnum \!tgCode=1 
        \ifx \!tgValue\empty 
          \the\ColumnWidthFactor
        \else
          \!tgValue    
        \fi
        \!taCWU        
      \else 
        \!tgValue      
      \fi}}}%
  \!ttemp
  a{\EndTableParBox}}

\def\!tgInsertKern{%
  \edef\!ttemp{%
    \kern
    \ifnum \!tgCode=1 
      \ifx \!tgValue\empty 
        \the\KernFactor
      \else
        \!tgValue    
      \fi
      \!taKU         
    \else 
      \!tgValue      
    \fi}%
  \edef\!ttemp{%
    \noexpand\ReadFormatKeys
    \ifh@            
      b{\!ttemp}
    \fi
    \ifv@            
      a{\!ttemp}
    \fi}%
  \!ttemp}

\def\NewFormatKey#1{%
  \!thx\def\!thx\!ttempa\!thx{\string #1}%
  \!thx\def\!thx\!ttempb\!thx{\csname !tk<\!ttempa>\endcsname}%
  \ifnum \TracingKeys>0
    \!tkReportNewKey
  \fi
  \!thx\ifx \!ttempb \relax
    \!thx\!tkDefineKey
  \else 
    \!thx\!tkRejectKey
  \fi}

\def\!tkReportNewKey{%
  \!taToksA\!thx{\!ttempa}%
  \!thMessage{NEW KEY: "\the\!taToksA"}}

\def\!tkDefineKey{%
  \!thx\def\!ttempb}%

\def\!tkRejectKey{%
    \!taToksA\!thx{\!ttempa}%
    \!thError{Key letter "\the\!taToksA" already used}
      {\!thReadErrorMsg\!tkFixIt}
    \def\!tkGarbage}%

\!thStoreErrorMsg\!tkFixIt{%
  You'd better type \space 'E' \space and fix your file.}

\def\ReadFormatKeys#1{%
  \!thx\def\!thx\!ttempa\!thx{\string #1}%
  \!thx\def\!thx\!ttempb\!thx{\csname !tk<\!ttempa>\endcsname}%
  \ifnum \TracingKeys>1
    \!tkReportKey
  \fi
  \!thx\ifx \!ttempb\relax 
    \!thx\!tkReplaceKey
  \else
    \!thx\!ttempb
  \fi}

\def\!tkReportKey{%
  \!taToksA\!thx{\!ttempa}%
  \!thMessage{KEY: "\the\!taToksA"}}

\def\!tkReplaceKey{%
  \!taToksA\!thx{\!ttempa}%
  \!thError {Undefined format key "\the\!taToksA"}
    {\!thReadErrorMsg\!tkUndefined ^^J\!thReadErrorMsg\!tkBadKey}
  \!tkReplaceKeyA}

\def\!tkReplaceKeyA{%
  \!thGetReplacement{\!thReadErrorMsg\!tkReplace}\!tkReplacement
  \!thx\ReadFormatKeys\!tkReplacement}

\!thStoreErrorMsg\!tkUndefined{%
  The format key in " "'s on the next to top line is undefined.}
\!thStoreErrorMsg\!tkBadKey{%
  Type \space E \space to quit now, or
  \space<CR> \space and respond to next prompt.}
\!thStoreErrorMsg\!tkReplace{%
  Type \space<replacement key><CR> \space,
   or simply \space<CR> \space to skip offending key:}

\NewFormatKey b#1{%
  \!thx\!tkJoin\!thx{\the\!taDataColumnTemplate}{#1}%
  \ReadFormatKeys}

\def\!tkJoin#1#2{%
  \!taDataColumnTemplate{#2#1}}%

\NewFormatKey a#1{%
  \!taDataColumnTemplate\!thx{\the\!taDataColumnTemplate #1}%
  \ReadFormatKeys}

\NewFormatKey \{{%
  \!taDataColumnTemplate=\!thx{\!thx{\the\!taDataColumnTemplate}}%
  \ReadFormatKeys}

\NewFormatKey *#1#2{%
  \!taCountA=#1\relax
  \!taToksA={}%
  \!thLoop 
    \ifnum \!taCountA > 0
    \!taToksA\!thx{\the\!taToksA #2}%
    \advance\!taCountA -1
  \repeat 
  \!thx\ReadFormatKeys\the\!taToksA}

\NewFormatKey \LeftGlue#1{%
  \!taLeftGlue{#1}%
  \ReadFormatKeys}

\NewFormatKey \RightGlue#1{%
  \!taRightGlue{#1}%
  \ReadFormatKeys}

\NewFormatKey c{%
  \ReadFormatKeys 
  \LeftGlue\hfil
  \RightGlue\hfil}

\NewFormatKey l{%
  \ReadFormatKeys 
  \LeftGlue{}   
  \RightGlue\hfil}

\NewFormatKey r{%
  \ReadFormatKeys 
  \LeftGlue\hfil
  \RightGlue{}}

\NewFormatKey k{%
  \h@true
  \v@true
  \!tgGetValue{\!tgInsertKern}}

\NewFormatKey i{%
  \h@true
  \v@false
  \!tgGetValue{\!tgInsertKern}}
  
\NewFormatKey j{%
  \h@false
  \v@true
  \!tgGetValue{\!tgInsertKern}}

\NewFormatKey n{%
  \def\!tnStyle{}%
   \futurelet\!tnext\!tnTestForBracket}

\NewFormatKey N{%
  \def\!tnStyle{$}%
   \futurelet\!tnext\!tnTestForBracket}

\NewFormatKey m{%
  \ReadFormatKeys b$ a$}

\NewFormatKey M{%
  \ReadFormatKeys \{ b{$\displaystyle} a$}

\NewFormatKey \m{%
  \ReadFormatKeys l b{{}} m}

\NewFormatKey \M{%
  \ReadFormatKeys l b{{}} M}

\NewFormatKey f#1{%
  \ReadFormatKeys b{#1}}

\NewFormatKey B{%
  \ReadFormatKeys f\bf}

\NewFormatKey I{%
  \ReadFormatKeys f\it}

\NewFormatKey S{%
  \ReadFormatKeys f\sl}

\NewFormatKey R{%
  \ReadFormatKeys f\rm}

\NewFormatKey T{%
  \ReadFormatKeys f\tt}

\NewFormatKey p{%
  \!tgGetValue{\!tgSetUpParBox}}

\NewFormatKey w{%
  \!tkTestForBeginFormat w{\!tgGetValue{\!tfSetWidth}}}

\NewFormatKey s{%
  \!taOnceOnlyTabskipfalse    
  \!tkTestForBeginFormat t{\!tgGetValue{\!tfSetTabskip}}}

\NewFormatKey o{%
  \!taOnceOnlyTabskiptrue
  \!tkTestForBeginFormat o{\!tgGetValue{\!tfSetTabskip}}}

\NewFormatKey |{%
  \!tkTestForBeginFormat |{\!tgGetValue{\!tfSetVrule}}}

\NewFormatKey \|{%
  \!tkTestForBeginFormat \|{\!tfSetAlternateVrule}}

\NewFormatKey .{%
  \!tkTestForBeginFormat.{\!tfFinishFormat}} 

\NewFormatKey \EndFormat{%
  \!tkTestForBeginFormat\EndFormat{\!tfFinishFormat}} 

\NewFormatKey ]{%
  \!tkTestForReFormat ] \!tfEndReFormat}

\def\!tkTestForBeginFormat#1#2{%
  \if!taBeginFormat  
    \def\!ttemp{#2}%
    \!thx \!ttemp    
  \else
    \toks0={#1}%
    \toks2=\!thx{\string\ReFormat}%
    \!thx \!tkImproperUse
  \fi}   

\def\!tkTestForReFormat#1#2{%
  \if!taBeginFormat  
    \toks0={#1}%
    \toks2=\!thx{\string\BeginFormat}%
    \!thx \!tkImproperUse
  \else
    \def\!ttemp{#2}%
    \!thx \!ttemp    
  \fi}   

\def\!tkImproperUse{%
  \!thError{\!thReadErrorMsg\!tkBadUseA "\the\toks0 "}%
    {\!thReadErrorMsg\!tkBadUseB \the\toks2 \space command.
    ^^J\!thReadErrorMsg\!tkBadKey}%
  \!tkReplaceKeyA}
 
\!thStoreErrorMsg\!tkBadUseA{Improper use of key }  
\!thStoreErrorMsg\!tkBadUseB{%
  The key mentioned above can't be used in a }

\def\!tnTestForBracket{%
  \ifx [\!tnext
    \!thx\!tnGetArgument
  \else
    \!thx\!tnGetCode
  \fi}

\def\!tnGetCode#1 {
  \!tnConvertCode #1..!}

\def\!tnConvertCode #1.#2.#3!{%
  \begingroup
    \aftergroup\edef \aftergroup\!ttemp \aftergroup{%
      \aftergroup[%
      \!taCountA #1
      \!thLoop
        \ifnum \!taCountA>0
        \advance\!taCountA -1
        \aftergroup0
      \repeat
      \def\!ttemp{#3}%
      \ifx\!ttemp \empty
      \else
        \aftergroup.
        \!taCountA #2
        \!thLoop 
          \ifnum \!taCountA>0
          \advance\!taCountA -1
          \aftergroup0
        \repeat
      \fi 
      \aftergroup]\aftergroup}%
    \endgroup\relax
    \!thx\!tnGetArgument\!ttemp}
  
\def\!tnGetArgument[#1]{%
  \!tnMakeNumericTemplate\!tnStyle#1..!}

\def\!tnMakeNumericTemplate#1#2.#3.#4!{
  \def\!ttemp{#4}%
  \ifx\!ttemp\empty
    \!taDimenC=0pt
  \else
    \setbox0=\hbox{\m@th #1.#3#1}%
    \!taDimenC=\wd0
  \fi
  \setbox0 =\hbox{\m@th #1#2#1}%
  \!thToksEdef\!taDataColumnTemplate={%
    \noexpand\!tnSetNumericItem
    {\the\wd0 }%
    {\the\!taDimenC}%
    {#1}%
    \the\!taDataColumnTemplate}  
  \ReadFormatKeys}

\def\!tnSetNumericItem #1#2#3#4 {
  \!tnSetNumericItemA {#1}{#2}{#3}#4..!}

\def\!tnSetNumericItemA #1#2#3#4.#5.#6!{%
  \def\!ttemp{#6}%
  \hbox to #1{\hss \m@th #3#4#3}%
  \hbox to #2{%
    \ifx\!ttemp\empty
    \else
       \m@th #3.#5#3%
    \fi
    \hss}}

\def\MakeStrut#1#2{%
  \vrule width0pt height #1 depth #2}

\def\StandardTableStrut{%
  \MakeStrut{\StrutHeightFactor\StrutUnit}
    {\StrutDepthFactor\StrutUnit}}

\def\AugmentedTableStrut#1#2{%
  \dimen@=\StrutHeightFactor\StrutUnit
  \advance\dimen@ #1\StrutUnit
  \dimen@ii=\StrutDepthFactor\StrutUnit
  \advance\dimen@ii #2\StrutUnit
  \MakeStrut{\dimen@}{\dimen@ii}}

\def\Enlarge#1#2{
  \!taDimenA=#1\relax
  \!taDimenB=#2\relax
  \let\!TsSpaceFactor=\empty
  \ifmmode
    \!thx \mathpalette
    \!thx \!TsEnlargeMath
  \else
    \!thx \!TsEnlargeOther
  \fi}

\def\!TsEnlargeOther#1{%
  \ifhmode
    \setbox\z@=\hbox{#1%
      \xdef\!TsSpaceFactor{\spacefactor=\the\spacefactor}}%
  \else
    \setbox\z@=\hbox{#1}%
  \fi
  \!TsFinishEnlarge}
    
\def\!TsEnlargeMath#1#2{%
  \setbox\z@=\hbox{$\m@th#1{#2}$}%
  \!TsFinishEnlarge}

\def\!TsFinishEnlarge{%
  \dimen@=\ht\z@
  \advance \dimen@ \!taDimenA
  \ht\z@=\dimen@
  \dimen@=\dp\z@
  \advance \dimen@ \!taDimenB
  \dp\z@=\dimen@
  \box\z@ \!TsSpaceFactor{}}

\def\OpenUp#1#2{%
  \advance \StrutHeightFactor #1\relax
  \advance \StrutDepthFactor #2\relax}

\def\BeginTable{%
  \futurelet\!tnext\!ttBeginTable}

\def\!ttBeginTable{%
  \ifx [\!tnext
    \def\!tnext{\!ttBeginTableA}%
  \else 
    \def\!tnext{\!ttBeginTableA[c]}%
  \fi
  \!tnext}

\def\!ttBeginTableA[#1]{%
  \if #1u
    \ifmmode                 
      \def\!ttEndTable{
        \relax}
    \else                   
      \bgroup
      \def\!ttEndTable{%
        \egroup}%
    \fi
  \else
    \hbox\bgroup $
    \def\!ttEndTable{%
      \egroup 
      $
      \egroup}
    \if #1t%
      \vtop
    \else
      \if #1b%
        \vbox
      \else
        \vcenter 
      \fi
    \fi
    \bgroup      
  \fi
  \advance\!taRecursionLevel 1 
  \let\!ttRightGlue=\relax  
  \everycr={}
  \ifnum \!taRecursionLevel=1
    \!ttInitializeTable
  \fi}

\bgroup
  \catcode`\|=13
  \catcode`\"=13
  \catcode`\~=13
  \gdef\!ttInitializeTable{%
    \let\!ttTie=~ 
    \let\!ttDH=\- 
    \catcode`\|=\active
    \catcode`\"=\active
    \catcode`\~=\active
    \def |{\unskip\!ttRightGlue&&}
    \def\|{\unskip\!ttRightGlue&\omit\!ttAlternateVrule}%
    \def"{\unskip\!ttRightGlue&\omit&}
    \def~{\kern .5em}
    \def\\{\!ttEndOfRow}%
    \def\-{\!ttShortHrule}%
    \def\={\!ttLongHrule}%
    \def\_{\!ttFullHrule}%
    \def\Left##1{##1\hfill\null}
    \def\Center##1{\hfill ##1\hfill\null}
    \def\Right##1{\hfill##1}%
    \the\EveryTable}
\egroup

\let\!ttRightGlue=\relax  

\def\!ttDoHalign{%
  \baselineskip=0pt \lineskiplimit=0pt \lineskip=0pt %
  \tabskip=0pt
  \halign \the\!taTableSpread \bgroup
   \span\the\!taPreamble
   \ifx \!tfRowOfWidths \empty
   \else 
     \!tfRowOfWidths \cr %
   \fi}

\def\EndTable{%
  \egroup 
  \!ttEndTable}

\def\!ttEndOfRow{%
  \futurelet\!tnext\!ttTestForBlank}

\def\!ttTestForBlank{%
  \ifx \!tnext\!thSpaceToken  
    \!thx\!ttDoStandard
  \else
    \!thx\!ttTestForZero
  \fi}
  
\def\!ttTestForZero{%
  \ifx 0\!tnext
    \!thx \!ttDoZero
  \else
    \!thx \!ttTestForPlus
  \fi}

\def\!ttTestForPlus{%
  \ifx +\!tnext
    \!thx \!ttDoPlus
  \else
    \!thx \!ttDoStandard
  \fi}

\def\!ttDoZero#1{
  \cr} 

\def\!ttDoPlus#1#2#3{
  \AugmentedTableStrut{#2}{#3}%
  \cr} 

\def\!ttDoStandard{%
  \StandardTableStrut
  \cr}

\def\!ttAlternateVrule{%
  \!tgGetValue{\!ttAVTestForCode}}  

\def\!ttAVTestForCode{%
  \ifnum \!tgCode=2              
    \!thx\!ttInsertVrule         
  \else
    \!thx\!ttAVTestForEmpty
  \fi}

\def\!ttAVTestForEmpty{%
  \ifx \!tgValue\empty           
    \!thx\!ttAVTestForBlank
  \else
    \!thx\!ttInsertVrule         
  \fi}

\def\!ttAVTestForBlank{%
  \ifx \!ttemp\!thSpaceToken     
    \!thx\!ttInsertVrule
  \else
    \!thx\!ttAVTestForStar 
  \fi}

\def\!ttAVTestForStar{%
  \ifx *\!ttemp                  
    \!thx\!ttInsertDefaultPR     
  \else
    \!thx\!ttGetPseudoVrule       
  \fi}

\def\!ttInsertVrule{%
  \hfil 
  \vrule \!thWidth
    \ifnum \!tgCode=1
      \ifx \!tgValue\empty 
        \LineThicknessFactor
      \else
        \!tgValue
      \fi
      \LineThicknessUnit
    \else
      \!tgValue
    \fi
  \hfil
  &}

\def\!ttInsertDefaultPR*{%
  \PseudoVrule    
  &}

\def\!ttGetPseudoVrule#1{%
  \toks0={#1}%
  #1&}

\def\PseudoVrule{}

\def\!ttuse#1{%
  \ifnum #1>\@ne 
    \omit 
    \mscount=#1 
    \advance\mscount by \m@ne
    \advance\mscount by \mscount
    \!thLoop 
      \ifnum\mscount>\@ne 
      \sp@n 
    \repeat 
    \span 
  \fi}

\def\!ttUse#1[{%
  \!ttuse{#1}%
  \ReFormat[}

\def\!ttFullHrule{%
  \noalign
  \bgroup
  \!tgGetValue{\!ttFullHruleA}}

\def\!ttFullHruleA{%
  \!ttGetHalfRuleThickness 
  \hrule \!thHeight \dimen0 \!thDepth \dimen0
  \penalty0 
  \egroup} 

\def\!ttShortHrule{%
  \omit
  \!tgGetValue{\!ttShortHruleA}}

\def\!ttShortHruleA{%
  \!ttGetHalfRuleThickness 
  \leaders \hrule \!thHeight \dimen0 \!thDepth \dimen0 \hfill
  \null    
  \ignorespaces} 

\def\!ttLongHrule{%
  \omit\span\omit\span \!ttShortHrule}

\def\!ttGetHalfRuleThickness{%
  \dimen0 =
    \ifnum \!tgCode=1
      \ifx \!tgValue\empty
        \LineThicknessFactor
      \else
        \!tgValue    
      \fi
      \LineThicknessUnit
    \else
      \!tgValue      
    \fi
  \divide\dimen0 2 }

\def\WidenTableBy#1{%
  \ifdim #1=0pt
    \!taTableSpread={}%
  \else
    \!taTableSpread={spread #1}%
  \fi}

\def\JustLeft{%
  \omit \let\!ttRightGlue=\hfill}
\def\JustCenter{%
  \omit \hfill\null \let\!ttRightGlue=\hfill}

\let\\=\!tacr
\catcode`\!=12
\catcode`\@=12

\documentstyle[12pt,fleqn,epsfig]{article} 
\def\stot{\sigma_{\rm tot}}

    \def\pbar{\mbox{$\bar {\rm p}$}}

\newcounter{list}
{\begin{list}
{(\Alph{list})}
{\usecounter{list}
\setlength{\rightmargin}{\leftmargin}}
}%
{\end{list}}
\newenvironment{alphList}%
{\begin{list}
{(\alph{list})}
{\usecounter{list}
\setlength{\rightmargin}{\leftmargin}}
}%
{\end{list}}
{\begin{list}
{(\Roman{list})}
{\usecounter{list}
\setlength{\rightmargin}{\leftmargin}}
}%
{\end{list}}
\newenvironment{romList}%
{\begin{list}
{(\roman{list})}
{\usecounter{list}
\setlength{\rightmargin}{\leftmargin}}
}%
{\end{list}}
{\begin{list}
{\arabic{list}.}
{\usecounter{list}
\setlength{\rightmargin}{\leftmargin}}
}%
{\end{list}}
\def\PseudoVrule{\hfil \vrule \hskip2pt \vrule \hfil} 
\begin{document}    
\begin{titlepage} 
%
\newcommand\reportnumber{490} 
\newcommand\mydate{September, 1995} 
\newlength{\nulogo} 
\settowidth{\nulogo}{\small\sf{N.U.H.E.P. Report No. \reportnumber}}
\title{
\vspace{-.8in} 
\hfill\fbox{{\parbox{\nulogo}{\small\sf{Northwestern University: \\
N.U.H.E.P. Report No. \reportnumber\\
          \mydate}}}}
          \vspace{.5in} \\
{The High Energy Behavior of the Forward Scattering Parameters---An Amplitude
Analysis Update
 }
\vspace{.2in}\\
}
\author{
M.~M.~Block
\thanks{Work partially supported by Department of Energy contract
DA-AC02-76-Er02289 Task B.}\vspace{-5pt}   \\
{\small\em Department of Physics and Astronomy,} \vspace{-5pt} \\ 
{\small\em Northwestern University, Evanston, IL 60208}\\
\vspace{-5pt}\\
%
B.~Margolis
\thanks{Deceased.
}\vspace{-5pt} \\
{\small\em  Physics Department,}\vspace{-5pt}  \\
{\small\em McGill University, Montreal, Canada H3A 2T8}\\
\vspace{-5pt}\\
%
A.~R.~White
\thanks{Work supported by
Department of Energy contract
W-31-109-ENG-38.}\vspace{-5pt} \\
{\small\em High Energy Physics Division,} \vspace{-5pt} \\
{\small\em Argonne National Laboratory, Argonne, Il 60439}  \\
\vspace{.2in}\\
%
{\small \sf Paper presented by Martin M. Block}\\[-4pt]
{\small \sf at the}\\[-4pt]
{\small \sf VIth Blois Workshop, Chateau de Blois, France, June, 1995}\\[-4pt]
}    
\date{} 
\maketitle
\vfill
\vspace{.2in}
\date{} 
%
%

\renewcommand\thepage{\ }
%
\begin{abstract} 
Utilizing the most recent experimental data, we reanalyze high energy \pbar p
and pp data, using the asymptotic amplitude
analysis, under the
assumption that we have reached `asymptopia'. This analysis gives strong
evidence for a $\log \,(s/s_0)$ dependence at {\em current} energies
and {\em not} $\log^2 (s/s_0)$, and also
demonstrates that odderons are {\em not} necessary to explain the
experimental data.

\end{abstract}  
\end{titlepage} 
%
\pagenumbering{arabic}
\renewcommand{\thepage}{-- \arabic{page}\ --}  
\renewcommand{\thesection}{\Roman{section}}  

\section{Dedication}
This work is dedicated to the memory of Bernard Margolis, Rutherford Professor
of Physics at McGill University, Montreal, Canada, who died shortly after this
paper was presented in June, 1995.  He was a magnificent physicist and a more
magnificent friend. He will be sorely missed!
\section{Asymptotic Amplitude Analysis}
In spite of the fact that there are excellent arguments
\cite{others}
that the
energy region in which present experiments are conducted---even at the
Tevatron Collider---is too low to be
considered asymptotic, we will consider here the consequences of assuming
the {\em opposite}. This allows us to test specific hypotheses
using a well-defined phenomenological
analysis .
We caution the reader that we
{\em don't believe} we are in `asymptopia' and thus {\em don't believe} the
analysis is applicable as a true asymptotic analysis. We {\em do believe}
that present day energies are too low to make  a truly asymptotic analysis.
Nonetheless, we feel that such an analysis is valuable as a guideline to
what is and is not happening at present energies.

We apply a ``standard'' asymptotic analytic amplitude
analysis
pro\-ce\-dure\cite{others} to the now-available data
on $\stot$, the total cross section and $\rho$, the ratio of the real
to the imaginary portion of the forward scattering amplitude, in the energy
region $\sqrt s$ = 5 to 1800
GeV, including the new CDF cross sections. The data are parameterized in terms
of even and odd analytic
amplitudes. Consistent with all asymptotic theorems, this allows use of
even amplitudes varying as fast as $\log^{2}(s/s_0)$ and odd
amplitudes (the `Odderon' family) that do {\em not} vanish as
$s\rightarrow\infty$.

We show here only the large $s$ limit of the even and odd
amplitudes that are used\cite{others}. We make five fits to the data:
\begin{romList}
\item Fit 1: $\log^{2}(s/s_0)$ energy dependence for the cross section, with
no  Odderon amplitude,
\item Fit 2: $\log^{2}(s/s_0)$ energy dependence for the cross section, with
an Odderon amplitude whose cross sectional dependence is $\log s$, the most
rapid behavior allowed by asymptotic theorems,
\item Fit 3: $\log^{2}(s/s_0)$ energy dependence for the cross section, with
an Odderon amplitude whose cross sectional dependence is constant,
\item Fit 4: $\log \,(s/s_0)$ energy dependence for the cross section, with
no  Odderon amplitude,
\item Fit 5: $\log \,(s/s_0)$ energy dependence for the cross section, with
an Odderon amplitude whose cross sectional dependence  is constant, the most
rapid behavior allowed by asymptotic theorems for this choice of even
amplitude.
\end{romList}

In all cases, an odd amplitude which vanishes with increasing energy is
also employed, as well as an even amplitude that mimics Regge behavior.
\subsection{log$\,{}^{2}(s)$ Energy Behavior}
We introduce $f_+$ and $f_-$, the even and odd (under crossing) analytic
amplitudes
at $t=0$, and define the $\bar {\rm p}$p and pp forward scattering
amplitudes by
$f_{\bar{\rm p}{\rm p}}= f_{+} + f_{-}\,\,\,\,{\rm and}\,\,\,\,
f_{\rm pp}=f_+ - f_-,$
giving total cross sections $\stot$ and
the $\rho $-values
\begin{eqnarray}
\sigma_{\bar{\rm p}{\rm p}}=\frac{4\pi}{p}
                    \,{\rm Im}\,f_{\bar{\rm p}{\rm p}}
,\,\,\,\,
\sigma_{{\rm pp}}=\frac{4\pi}{p}\,{\rm Im}\,f_{\rm pp},\,\,\,\,
\rho_{\bar{\rm p}{\rm p}}=
\frac{ {\rm Re}\,f_{\bar{\rm p}{\rm p}} }
{ {\rm Im}\,f_{ \bar{\rm p}{\rm p} } }
,\,\,\,\, {\rm and}\,\,\,\,
\rho_{{\rm pp}}=
\frac{ {\rm Re}\,f_{{\rm pp}} }
{ {\rm Im}\,f_{{\rm pp}} }.
\end{eqnarray}
We parameterize the `conventional' even and odd amplitudes $f_+$ and $f_-$
by :
\begin{eqnarray}
     \frac{4\pi}{p}f_+ &=& i \left ( A+
     \beta \left [\log \left (\frac{s}{s_0}\right ) - i \frac{\pi}{2}
     \right ]^2 + c\,s^{\mu -1}e^{i\pi(1-\mu )/2}\right ) \label{eq:even2} \\
     \frac{4\pi}{p}f_- &=& -Ds^{\alpha -1} e^{i\pi (1-\alpha )/2}.
\label{eq:odd}
\end{eqnarray}

The parameter $\alpha $ in Eq~(\ref{eq:odd}) turns out to be about 0.5,
and thus
this odd amplitude vanishes as $s\rightarrow\infty$.

Asymptotic theorems by Eden and Kinoshita\cite{others} prove
that the {\em difference} of cross sections can not grow
faster than $\log^{\gamma /2}(s)$, when the cross section grows as
$\log^{\gamma }(s)$. Thus, odd amplitudes which do {\em not}
vanish as
$s\rightarrow\infty$ for this case are :

     $\frac{4\pi}{p}f_-^{(0)}= -\,\epsilon ^{(0)},\,\,\,\,
     \frac{4\pi}{p}f_-^{(1)} =-\left [
                      \log \left ( \frac {s}{s_0} \right ) -i \frac{\pi}{2}
                       \right ] \epsilon ^{(1)},\,\,\,\, {\rm and},\,\,\,\,
     \frac{4\pi}{p}f_-^{(2)} = -\left [
                      \log \left ( \frac {s}{s_0} \right ) -i \frac{\pi}{2}
                      \right ]^{2} \epsilon ^{(2)}.$

The complete odd amplitude is formed by adding any one (or none)
of the $f^{(i)}_-$  to the conventional odd amplitude $f_-$
of Eq~(\ref{eq:odd}).  We then fit the experimental $\rho $ and
$\stot $ data, for
both pp and $\bar{\rm p}$p, for energies between 5 and 1800 GeV, to obtain
the real constants $A,\beta ,s_0,c,\mu ,D,\alpha ,\epsilon ^{(i)}$.  The data
used below 500 GeV are listed in \cite{others}, and the high energy
points are from UA1, UA4, E710
and CDF\cite{others}.
We emphasize that what we really fit for the UA4 and CDF cross sections is
the measured
experimental quantity $\stot\times (1+\rho^2)$, which is appropriate for
experiments that measure a `luminosity-free' cross section, whereas for UA1
and the 1020 GeV point of E710,
we fit the experimental quantity $\stot\times \sqrt {1+\rho^2}$,
which was
their experimentally measured quantity (they measured a
`luminosity-dependent' cross section).

\subsection{Fitted Results for log$\,{}^{2}(s)$ Behavior}
\begin{figure}[htb]
\centerline{\psfig{figure=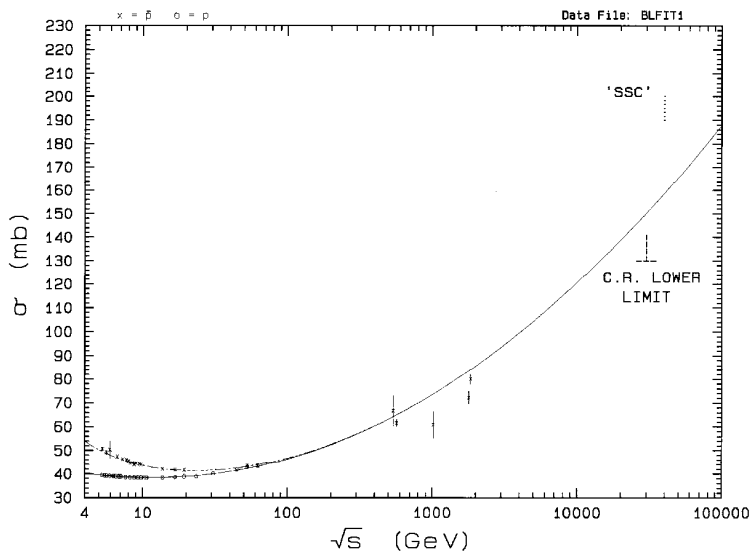,width=4.25in}}
\caption{\protect{\footnotesize {The total cross section  $\sigma_{tot}$, in
mb, for
$\bar {\rm{p}}$p and
pp scattering {\it vs.} the energy, $\protect\sqrt s$, in GeV, for Fit 1,
described
in Table I.  The fit was made with a $\log^2(s)$ energy variation, and no
Odderon.  The crosses are for the $\bar {\rm{p}}$p experimental data and
the circles
indicate pp data. The dot-dashed curves are for $\bar {\rm{p}}$p, and the
solid curves for pp. The pp cosmic-ray lower limit\protect\cite{others} is
appended to
the curve, but is {\em not} used in the fit.}}}\protect\label{sfit1}
\end{figure}

\begin{figure}[hbt]
\centerline{\psfig{figure=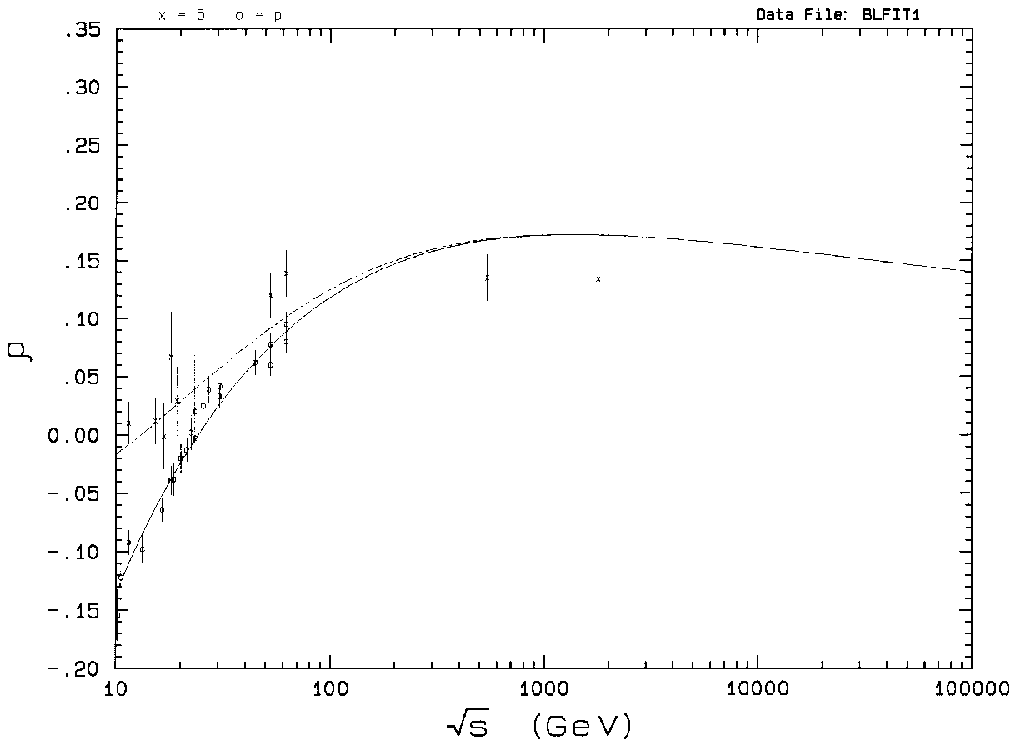,width=4.25in}}
\caption{\protect{\footnotesize {The $\rho$-value for $\bar {\rm{p}}$p and
pp scattering {\it vs.} the energy, $\protect\sqrt s$, in GeV, for Fit 1,
described
in Table I.  The fit was made with a $\log^2(s)$ energy variation, and no
Odderon. The crosses are for the $\bar {\rm{p}}$p experimental data and
the circles
indicate pp data. The dot-dashed curves are for $\bar {\rm{p}}$p, and the
solid curves for pp.}}}\protect\label{rfit1}
\end{figure}

\begin{romList}
\item Fit 1---This fit uses no Odderons in the odd amplitude and uses the
even amplitude of Eq~(\ref{eq:even2}).
The $\chi^2$/d.f. ($\chi^2$/degree of freedom)
for the fit
was 1.94, a rather large number.  The fitted constants are shown in
Table \ref{ta:amp},
Fit 1---the computed
curves are shown in Fig.~\ref{sfit1} (for $\stot$) and Fig.~\ref{rfit1} (for
$\rho$).
\noindent The most obvious features of the fit are
\begin{alphList}
 \item the predicted value
of the total cross section is much
too high
to fit the experimental values (E710 and CDF) at 1800 GeV,
\item it predicts much too high a $\rho $-value at 546 GeV.
\end{alphList}

We conclude that a simple $\log^2(s)$ fit does not fit the data.

\item Fit 2---
We fit the data with an additional degree of
freedom, by adding Odderon 2  to $f_-$ of
Eq~(\ref{eq:odd}), along with the even
amplitude of Eq~(\ref{eq:even2}).  The parameters are summarized as Fit 2,
in Table \ref{ta:amp}. Again, we
conclude that this combination doesn't fit the data, since the high energy
cross section predicted at 1800 GeV is much too high.
Although the
$\rho$-value predicted at 540 GeV is slightly lower, the $\rho$ values
predicted are still too high.
\item Fit 3---The odd
amplitude added to the conventional $f_-$ of Eq~(\ref{eq:odd}) was
Odderon 1.  The parameters are given as Fit 3 in
Table \ref{ta:amp}.  Again, the fit suffers
from the
same defect as the Odderon 2 fit, giving much too high a total cross section
at 1800 GeV, as well as predicting a UA4/2
$\rho$-value which was much too high.

The addition of Odderon 0 can have no effect on
the cross section. Since it turns out to have a negligible effect on $\rho$,
we will not
consider it further.
\end{romList}
We conclude that an even amplitude varying as
$\log^2(s/s_0)$ {\em does not fit} the cross section data. We see that
the experimental cross section does not rise as rapidly as $\log^2(s/s_0)$,
in the present-day energy region. The addition
of an Odderon term does not change this conclusion.
\begin{figure}[htb]
\centerline{\psfig{figure=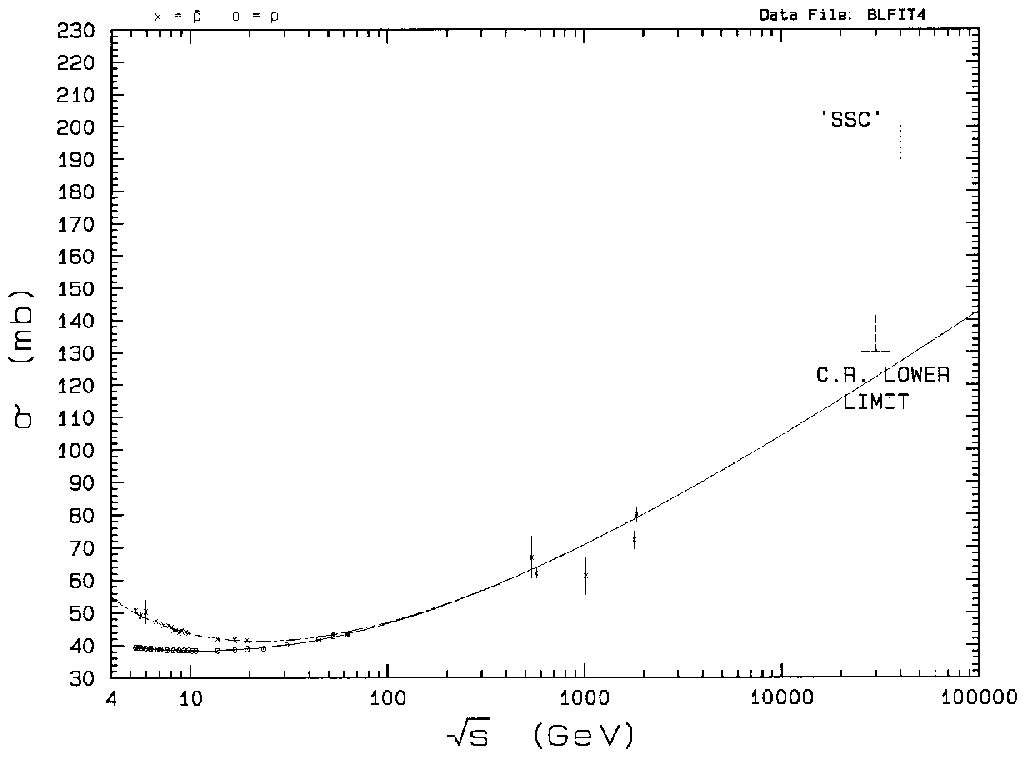,width=4.25in}}
\caption{\protect{\footnotesize {The total cross section  $\sigma_{tot}$, in
mb, for
$\bar {\rm{p}}$p and
pp scattering {\it vs.} the energy, $\protect\sqrt s$, in GeV, for Fit 4,
described
in Table I.  The fit was made with a $\log(s)$ energy variation, and no
Odderon.  The crosses are for the $\bar {\rm{p}}$p experimental data and
the circles
indicate pp data. The dot-dashed curves are for $\bar {\rm{p}}$p, and the
solid curves for pp. The pp cosmic-ray lower limit\protect\cite{others} is
appended to
the curve, but is {\em not} used in the fit.}}}\protect\label{sfit4}
\end{figure}
\begin{figure}[hbt]
\centerline{\psfig{figure=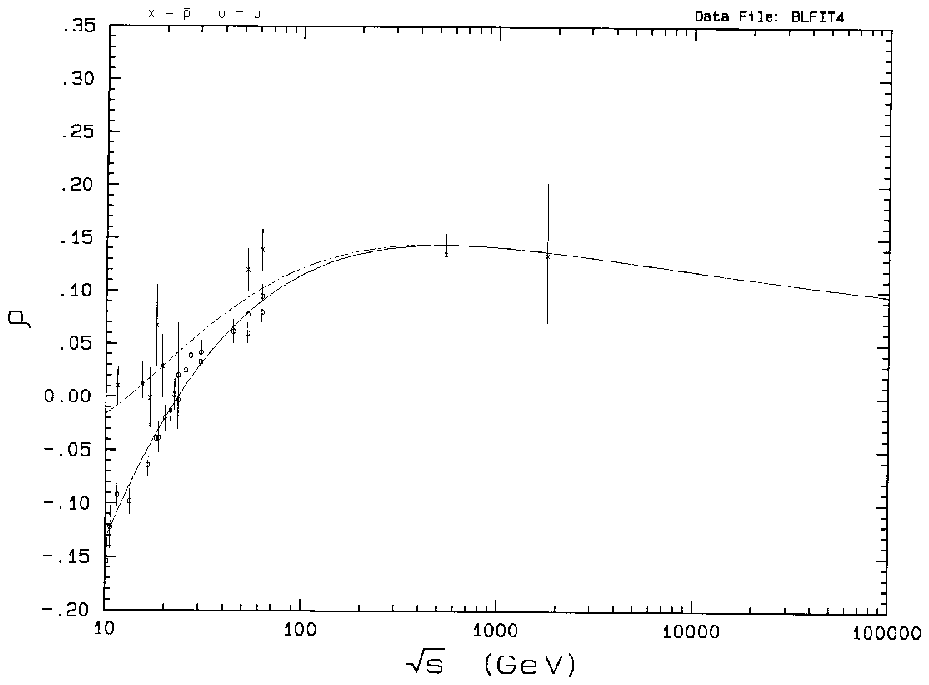,width=4.25in}}
\caption{\protect{\footnotesize {The $\rho$-value for $\bar {\rm{p}}$p and
pp scattering {\it vs.} the energy, $\protect\sqrt s$, in GeV, for Fit 4,
described
in Table I.  The fit was made with a $\log(s)$ energy variation, and no
Odderon. The crosses are for the $\bar {\rm{p}}$p experimental data and
the circles
indicate pp data. The dot-dashed curves are for $\bar {\rm{p}}$p, and the
solid curves for pp.}}}\protect\label{rfit4}
\end{figure}
\begin{table}[h,t]                   
%
\def\arraystretch{1.5}            
\begin{tabular}[b]{|l||l|l|l||l|l|}
     \cline{2-6}
      \multicolumn{1}{c|}{}
      &\multicolumn{3}{c||}{$\stot \sim \log^2(s/s_0)$}
      &\multicolumn{2}{c|}{$\stot \sim \log(s/s_0)$}\\
      \hline
      Parameters&Fit 1&Fit 2&Fit 3&Fit 4&Fit 5 \\ \hline
     $A$ (mb)&$40.3\pm .20$&$40.2\pm$ .24&$41.6\pm$ .04&$-26.0\pm 12.7$
     &$-22.5\pm 11.7$\\
     $\beta$ (mb)&$.47\pm .02$&$.48\pm .02$&$.57\pm .01$&$9.6\pm .9$
     &$9.4\pm .8$ \\
     $s_0$ (${\rm (GeV)}^2$)&$200\pm 20$&$207\pm 25$&$346\pm 11$&$500$
     &$500$ \\
     $D$ (mb${\rm (GeV)}^{2(1-\alpha)}$)&$-40.9\pm 1.9$&$-38.8\pm 2.1$
     &$-36.8\pm 1.6$&$-43.2\pm 2.1$&$-43.3\pm2.0$ \\
     $\alpha$&$.46\pm .02$&$.47\pm .02$&$.49\pm .02$&$.45\pm .02$
     &$.45\pm .02$ \\
     $c$ (mb${\rm (GeV)}^{2(1-\mu)})$&$30.9\pm 4.1$&$26.6\pm 4.3$
     &$5.9\pm 2.6$&$159\pm 17$&$154\pm 16$ \\
     $\mu$&$.46$&$.49$&$.49$&$.86\pm .01$&$.86\pm .01$ \\
     $\epsilon^{(2)}$ (mb)&&$-.024 \pm.011$&&& \\
     $\epsilon^{(1)}$ (mb)&&&$.035 \pm.040$&&$-.016\pm .043$ \\
     \hline
     $\chi^2/$d.f.&1.94&1.83&2.58&1.22&1.22 \\
     d.f.&82&81&81&82&81\\
     \hline
\end{tabular}
     \caption{\protect\small Results of fits to total cross sections and
     $\rho $-values, including Odderons. Fit 1, Fit 2 and Fit 3 correspond
     to an asymptotic
     cross section variation of $\log ^2(s/s_0)$, with no Odderon,
     Odderon~2 and
     Odderon~1, respectively, whereas Fit 4 and Fit 5 correspond
     to an energy dependence of $\log (s/s_0)$, with no Odderon and Odderon~1,
     respectively.\label{ta:amp}}
\end{table}
\subsection{log$\,(s)$ Energy Behavior}
Since the experimental cross section in the energy region 5-1800 GeV did
not vary as fast as $\log ^2(s/s_0)$, we now consider an asymptotic variation
that goes as $\log \,(s/s_0)$.  We substitute for the even
amplitude in
Eq~(\ref{eq:even2}) a new amplitude $f_+$ varying as $\log \,(s/s_0)$,
$
     \frac{4\pi}{p}f_+ = i \left (A+
     \beta\left [\log \left (\frac{s}{s_0}\right ) - i \frac{\pi}{2}
     \right ] + c\,s^{\mu -1}e^{i\pi(1-\mu )/2} \right ).
$
We use the conventional odd amplitude of Eq~(\ref{eq:odd}), along with no
Odderon or Odderon 1, in Fits 4 and 5, respectively. We make the
important observation that since
the energy variation of the cross section is now only $\log (s)$,
Odderon 2 is {\em not} allowed by the asymptotic theorems.

\subsection{Fitted Results for log$\,(s)$ Behavior}
\begin{romList}
\item
Fit 4---The data are fitted with a log$\,(s/s_0)$ cross section
energy behavior,
with no Odderon. The results are detailed in Table \ref{ta:amp}, and plotted
in Fig.~\ref{sfit4} and
Fig.~\ref{rfit4}. The fit is quite satisfactory, giving a $\chi^2$/d.f.
of 1.22, fitting reasonably well to all cross section data over the
entire range of energy.  Most importantly, it now fits the UA4/2 $\rho$-value
at 546 GeV,
as well as the E710 $\rho$-value at 1800 GeV.

\item Fit 5---
The data are fitted with a log$\,(s)$ cross section energy behavior,
along with Odderon 1. The results are given in Table
2. This fit (as is Fit 4) is
quite satisfactory, giving a $\chi^2$/d.f. of 1.24. Indeed, it is almost
indistinguishable from fit 4.
\end{romList}
We find that the experimental cross sections and $\rho$-values in the
energy domain 5--1800 GeV can be reproduced using a
$\log (s/s_0)$ energy variation. Further, the introduction of an Odderon
amplitude is not needed to explain the experimental data. Also, using the new
CDF cross
sections does not change these conclusions.

\end{document}